\documentstyle[12pt,axodraw,epsfig]{article}

\newlength{\dinwidth}
\newlength{\dinmargin}
\def\lapproxeq{\lower .7ex\hbox{$\;\stackrel{\textstyle
<}{\sim}\;$}}
\def\gapproxeq{\lower .7ex\hbox{$\;\stackrel{\textstyle
>}{\sim}\;$}}

\def\beq{\begin{equation}}
\def\eeq{\end{equation}}
\def\bea{\begin{eqnarray}}
\def\eea{\end{eqnarray}}

\def\GeV{\rm GeV}

\begin{document}
\titlepage
\begin{flushright}
Cavendish-HEP-2005/01 \\
January 2005 \\

\end{flushright}

\vspace*{0.5cm}

\begin{center}
{\Large \bf Gluon Distributions and Fits using Dipole Cross-Sections.}

\vspace*{1cm}
\textsc{R.S. Thorne\footnote{Royal Society University Research Fellow.}} \\

\vspace*{0.5cm} 
Cavendish Laboratory, University of Cambridge, \\ Madingley Road,
Cambridge, CB3 0HE, UK
\end{center}

\vspace*{0.5cm}

\begin{abstract}
I investigate the relationship between the gluon distribution obtained 
using a dipole model fit to low-$x$ data on $F_2(x,Q^2)$ and standard gluons 
obtained from global fits with 
the collinear factorization theorem at fixed order. I stress the necessity to 
do fits of this type carefully, and in particular to include the 
contribution from heavy flavours to the inclusive structure function. 
I find that the dipole cross-section must be rather
steeper than the gluon distribution, which at least partially explains why 
dipole model fits produce dipole cross-sections growing quite strongly 
at small $x$, while 
DGLAP based fits have valence-like, or even negative, small-$x$ gluons as 
inputs. However, I also find that the gluon distributions obtained
from the dipole fits are much 
too small to match onto the conventional DGLAP gluons at 
high $Q^2 \sim 50\GeV^2$, where the two approaches should coincide. The main
reason for this discrepancy is found to be the large approximations made 
in converting
the dipole cross-sections into structure functions using formulae which are
designed only for asymptotically small $x$. The shortcomings in this step 
affect the accuracy of the extracted dipole cross-sections in terms of size
and shape, and hence also in terms of interpretation, at all scales.     
\end{abstract}

\newpage

\section{Introduction}

\vspace{0.5cm}

In the description of structure function data the most conventional approach
used is the collinear factorization theorem, where total cross-sections 
are determined in terms of parton distributions and hard parton cross-sections
up to corrections of ${\cal O}(\Lambda_{QCD}^2/Q^2)$, i.e. higher-twist
corrections. The most complete method is to perform a so-called global fit 
\cite{MRST2001,CTEQ6} to all
data sensitive to parton distributions, so that the consistency of the fit to
a variety of different data sets is guaranteed. This is currently done at 
either NLO or NNLO in the strong coupling $\alpha_S$ and appears to work very 
well. However, there are some indications \cite{MRSTerror2} 
that the 
procedure is a little unreliable at small values of $x$ where a resummation
of large $\ln(1/x)$ terms may be important \cite{BFKL}, 
and by definition this whole 
approach fails at low values of $Q^2$ (where low appears to be somewhere
from $0.5$ to $4\,\GeV^2$). 

An alternative approach which circumvents the problem of low 
$Q^2$ and  
is particularly applicable to small $x$ is the colour dipole approach
\cite{Frankfurtcd,Muellercd,Nikolaevcd1,Nikolaevcd2}. Recently there have 
been a variety of fits, or at least comparisons, to small-$x$ structure 
function data using the dipole 
picture \cite{GBW1}-\cite{Forshaw1}. 
In this the free parameters of the fit are all mainly associated 
with the dipole cross-section, which it is very difficult to calculate from 
first principles but which may be modelled, with varying degrees, and 
different types of theoretical justification. If one also wants a finite 
photoproduction cross-section, a non-zero value must be chosen for the 
light-quark masses, appearing as a parameter in the dipole wavefunction 
which is calculated at LO, i.e. zeroth order in $\alpha_S$. It must be 
noted that
in order to apply the approach to very low $Q^2$ one must 
assume that perturbation theory is valid and higher order
QCD corrections to e.g. the photon wavefunction are meaningful and under 
control in this limit. This is yet to be proved. With this caveat in mind
it is true that a variety of approaches to modelling the dipole cross-section 
can be made to match data very well.

Even though there is no essential connection, 
the dipole cross-section approach is 
often linked to, and used together with, the approaches which deal with parton 
saturation at small $x$. 
It is commonly believed that the complications of small $x$ and low $Q^2$ are
entwined, with the assumed large parton distributions at small-$x$ leading
to significant reduction of the evolution due to 
the mixing of leading-twist parton evolution with higher-twist
multi-parton operators at low $Q^2$ \cite{MQ,LR} (though it is fair to say 
that the values of $x$ and $Q^2$ which are relevant are not so commonly 
agreed). There has recently been a great deal of work attempting, as far as 
is possible, to calculate the dipole cross-sections within this framework
of large densities and saturation (see e.g. \cite{BKetc}, or for a slightly 
different viewpoint \cite{classdip}), 
and many of the dipole fits, including 
some of the most successful, are based on these ideas. In some quarters
this has led to very strong claims that saturation has been discovered.
However, there is a conundrum. This picture of steeply growing parton 
distributions at small $x$ and low $Q^2$  
tamed by saturation is in conflict with the conventional DGLAP fits
which actually result in a small or even negative
gluon (and consequently $F_L(x,Q^2)$) at small $x$ and $Q^2$. It is essential 
to understand this before making any strong claims for saturation.   

In this paper I will investigate the cause of this inconsistency. I
will base this investigation very much on the completely 
standard assumption that
the QCD factorization theory is completely reliable, correct and 
quantitative at fairly high $Q^2$ (as long as $x$ is not too small). 
Hence the parton distributions obtained from global fits must be 
quantitatively correct in this region. I will use the known relationships 
between the dipole cross-sections and standard parton distributions to work
back from a fit to data performed in the dipole framework to obtain the
corresponding partons. First, I will
examine the question of whether a large/steep dipole cross-section 
actually means a large/steep gluon distribution, finding that the dipole 
cross-section is always steeper at small $x$ than its corresponding gluon
distribution. This partially explains the differing conclusions obtained from 
the DGLAP and dipole approaches, but is not the whole story. 
In order to investigate the consistency of the two approaches 
I obtain a gluon distribution which evolves in the same way as a DGLAP gluon
(at least for reasonably high $Q^2$) from a dipole model type fit to 
structure function data. I then make a 
comparison between the gluon distribution obtained at fairly high $Q^2$
from this dipole model fit and the gluon from a standard set of 
parton distributions. 
This gives strong evidence as to whether dipole motivated fits are 
truly quantitative, and whether the results from these fits are
to be taken seriously in detail.\footnote{The results of the fit to data 
using my particular model for the
dipole cross-section, or equivalently gluon distribution, could also be 
thought of as providing  some evidence as to whether or not 
saturation effects are 
important when using a dipole model fit. 
However, since the whole purpose of the paper is to question
the validity of this approach as far as any strong conclusions are concerned,
the implications from my particular
model as to the degree of saturation are really only a side issue.} 
I find that the comparison between 
the two illustrates a serious discrepancy, and point out that the reason 
for this discrepancy is the approximation inherent in the LO dipole 
wavefunctions.  I conclude that this result casts doubt on whether 
we should indeed treat the results of fits to HERA data using the dipole 
picture as telling us anything truly quantitative, and I explain
my reservations. Improvements in the quantitative form of the gluon 
distributions obtained from dipole model inspired fits rely mainly on 
increasing the precision of the wavefunctions used to obtain the structure 
functions from the dipole cross-section.

\vspace{0.5cm}

\section{The Relationship Between The Dipole Cross-Section and The Gluon 
Distribution}

\vspace{0.5cm}

The relationship between the gluon distribution and the dipole cross-section
was essentially worked out as soon as the dipole approach was proposed
\cite{Nikolaevcd1}, but it is nicely discussed in a pedagogical manner in  
\cite{Bialas} which explicitly shows the relationship between the dipole 
picture and the $k_T$-factorization theorem \cite{kttheorem} 
at LO, and also shows how this 
relationship breaks down beyond LO. My discussion partly follows this paper. 
The diagrams contributing to deep inelastic scattering at LO are shown below,
where the incoming gluons have finite transverse momentum $\vec k$.

\begin{picture}(600,200)(0,0)
\SetColor{Blue}
\SetScale{0.8}
\Photon(20,150)(75,150){5}{6}
\Photon(215,150)(270,150){5}{6}
\ArrowArc(145,110)(80,30,150)
\ArrowArc(145,190)(80,210,330)
\Gluon(120,40)(120,114){5}{3}
\Gluon(170,40)(170,114){5}{3}
\Text(53,137)[]{$1\!-\!z,\vec p$}
\Text(53,104)[]{$z,-\vec p$}
\Text(80,38)[]{$x,\vec k$}
\Text(148,38)[]{$x,\vec k$}
\Photon(320,150)(375,150){5}{6}
\Photon(515,150)(570,150){5}{6}
\ArrowArc(445,110)(80,30,150)
\ArrowArc(445,190)(80,210,330)
\Gluon(420,40)(420,114){5}{3}
\Gluon(470,40)(470,186){5}{6}
\Text(294,137)[]{$1\!-\!z,\vec p$}
\Text(294,104)[]{$z,-\vec p$}
\Text(319,38)[]{$x,\vec k$}
\Text(387,38)[]{$x,\vec k$}
\end{picture}

Within the LO $k_T$-factorization theory we can write, for example, the 
longitudinal $\gamma^{\star}p$ cross-section as 
\beq
\sigma_L(x,Q^2) \propto \int_0^1 dz[z(1-z)]^2\int \frac{d^2 k}{k^4} \int d^2
p \biggl( \frac{1}{\hat Q^2+p^2}-\frac{1}{\hat Q^2+(p+k)^2}\biggr)^2
f(x_g,k^2)\label{eq:xsection}
\eeq
where $f(x_g,k^2)$ is the unintegrated gluon distribution and $\hat Q^2
=z(1-z)Q^2$. A similar result, but slightly more complicated formula,
also holds for $\sigma_T$. 
Staying at strictly LO in $\ln(1/x)$ in the $k_T$-factorization theory, 
we work in the limit $x \to 0$, i.e. since 
\beq
\ln(x)=\ln(x_g) + \ln\bigl(\hat Q^2/(\hat Q^2 +\hat k^2 +(p')^2)\bigr),
\eeq
where $\vec p'=\vec p -(1-z)\vec p$, we simply make the identity $x=x_g$. 
In this limit Eq.(\ref{eq:xsection}) can be 
simplified significantly. Integrating over $z$ and $p$, which in this limit 
does not involve $f(x,k^2)$ we have the standard    
$k_T$-factorization expression, which can be written in terms of the 
structure function as 
\beq
F_L(x,Q^2) =  \int\frac{d k^2}{k^2} \,\frac{\alpha_S 2N_f}{6\pi}
\,\, h_L(k^2/Q^2) f(x,k^2).
\eeq 
Taking the double Mellin transformation $\int 
d Q^2 Q^{2-2\gamma}$
and $\int dx \, x^{N}$ we have the familiar expression.

\beq
\tilde F_L(N,\gamma) = \frac{\alpha_S 2N_f}{6\pi}\,
\tilde h_L(\gamma) \tilde f(N,\gamma)/\gamma
\equiv \frac{\alpha_S 2N_f}{6\pi}\, \tilde h_L(\gamma)\tilde g(N,\gamma),
\label{eq:LOkt}
\eeq
where $g(x,Q^2)=\int_0^{Q^2}\frac{d k^2}{k^2} f(x,k^2)$ is the 
integrated gluon distribution, and $\tilde h_L(\gamma)$ is the longitudinal 
impact factor first calculated in \cite{CatHaut}. 
An exactly analogous expression can be calculated for $F_2(x,Q^2)$, though
it is usually expressed in terms of $dF_2/d\ln Q^2$ in order to preserve 
finiteness in the infrared limit, i.e. 
\beq
\frac{d F_2(x,Q^2)}{d \ln Q^2} =  \int\frac{d k^2}{k^2} \,
\frac{\alpha_S 2N_f}{6\pi}
\,\, h_2(k^2/Q^2) f(x,k^2).
\eeq   

If the gluon distribution can be expressed in the simple form 
$g(N,Q^2) \sim (Q^2)^{\gamma(\alpha_S,N)}$ this leads to 
\beq
F_i(N,Q^2) =\frac{\alpha_S 2N_f}{6\pi} h_i(\gamma(\alpha_S,N))g(N,Q^2)
\label{eq:LOktsimp}
\eeq
which taking the inverse Mellin transformation becomes 
\beq
F(x,Q^2) = \frac{\alpha_S 2N_f}{6\pi} 
h(\gamma(\alpha_S,\ln(1/x)))\otimes g(x,Q^2).
\eeq
This is the standard result of the LO in $\ln(1/x)$ $k_T$-factorization 
theorem. Contrary to what seems to be common belief, the $k_T$-factorization
theorem is well defined beyond this order. Indeed, in \cite{CatHaut} it is 
demonstrated that $k_T$-factorization may be thought of as simply a 
reordering of the calculations performed within the collinear factorization 
theorem, and as such it is as well defined as collinear factorization, i.e. 
to all orders at leading twist.  

There is an alternative way to proceed from the starting point of 
Eq.(\ref{eq:xsection}). By using the identity
\beq
\frac{1}{\hat Q^2+p^2} = \frac{1}{2\pi}\int d^2r \exp(i p\cdot r) 
K_0(\hat Q r)
\eeq
and integrating over $p^2$, using the independence of $x_g$ on 
$p^2$ in the $x\to 0$ limit, 
one can equivalently write
\beq
\sigma = \frac{4\pi^2}{3}\int_0^1 dz\int d^2r \vert \Psi(r,z,Q)\vert^2 
\int\frac{d k^2}{k^4} \alpha_S f(x,k^2)(1-J_0(kr)).\label{eq:xsectiondip}
\eeq
$\vert \Psi(r,z,Q)\vert^2$ is the probability for a photon of virtuality 
$Q^2$ to fluctuate into a a dipole pair, as calculated in \cite{Nikolaevcd1},
and is explicitly
\bea
\vert \Psi_T(r,z,Q)\vert^2 &=& \frac{6\alpha}{4\pi^2}\sum_f e_f^2\bigl(
[z^2+(1-z)^2]\epsilon^2 K_1^2(\epsilon r) +m_f^2 K_0^2(\epsilon r)\bigr)\\
\vert \Psi_L(r,z,Q)\vert^2 &=& \frac{6\alpha}{\pi^2}\sum_f e_f^2\bigl(
Q^2[z^2+(1-z)^2]K_0^2(\epsilon r)\bigr),
\eea
where $\epsilon^2=z(1-z)Q^2 +m_f^2$, and $m_f$ is the mass of a given quark 
flavour. Hence, Eq.(\ref{eq:xsectiondip}) can be interpreted as  
\beq
\sigma = \int_0^1 dz\int d^2r \vert \Psi(r,z,Q)\vert^2 
\hat \sigma(x,r^2),\label{eq:xsectiondipa}
\eeq
where 
\beq
\hat \sigma(x,r^2) = \frac{4\pi^2}{3} 
\int\frac{d k^2}{k^4} \alpha_S f(x,k^2)(1-J_0(kr))\label{eq:dipsection}
\eeq 
may be associated with the dipole-proton cross-section. In the LO $\ln(1/x)$
limit this and Eq.(\ref{eq:LOkt}) are really equivalent, but because in
Eq.(\ref{eq:LOkt}) we have reference to the gluon density, which we think of 
as evolving perturbatively, and have an explicit factor of $\alpha_S$,
we think of this equation only having validity for $Q^2 \gg \Lambda_{QCD}^2$.
In principle the same issues exist for Eq.(\ref{eq:xsectiondipa}), with
$\hat \sigma(x,r^2)$ depending on both the (unintegrated) gluon distribution
and $\alpha_s$, as seen in Eq.(\ref{eq:dipsection}). However, ignoring these 
complications and proposing models for $\hat \sigma(x,r^2)$ valid for all
$r$, Eq.(\ref{eq:xsectiondipa}) is often used down to the photoproduction limit
of $Q^2=0$, albeit requiring regularization from finite light quark masses. 
However, as discussed in \cite{Bialas} the form of the expression in 
Eq.(\ref{eq:xsectiondipa}) is definitely not preserved beyond the 
leading $\ln(1/x)$ limit, with inclusion of real gluon kinematics spoiling
the diagonalization in the transverse size $\vec r$ of the incoming and 
outgoing dipoles.       

At LO we can investigate what the equivalence of 
the two approaches tells us. In the standard $k_T$-factorization theorem 
approach $F_i(N,Q^2)$ is given by Eq.(\ref{eq:LOktsimp}), or its equivalent 
for
$d\,F_2(N,Q^2)/d\ln Q^2$. Taking the Mellin transformation with respect to $x$ 
of the intermediate expression Eq.(\ref{eq:xsectiondip}) and using the 
equivalence we obtain
\beq
F_i(N,Q^2) =\frac{\alpha_S 2N_f}{6\pi} h_{id}
(\gamma(\alpha_S,N)) h_{dg}(\gamma(\alpha_S,N)) 
g(N,Q^2) 
\equiv 
\frac{\alpha_S 2N_f}{6\pi} h_i(\gamma(\alpha_S,N))g(N,Q^2),\label{eq:LOequiv}
\eeq
where $h_{id}(\gamma)$ comes from the probability of the photon splitting 
to the dipole, while $h_{dg}(\gamma)$ comes from the relationship between 
the dipole cross-section and the gluon distribution in 
Eq.(\ref{eq:dipsection}). 
Therefore, the effective coefficient function for the hard cross-section
$h_{i}(\gamma(\alpha_S,N))$ can be interpreted as 
the product of a photon-dipole coefficient
function $h_{id}(\gamma(\alpha_S,N))$ and a dipole-gluon coefficient 
function $h_{dg}(\gamma(\alpha_S,N))$, both of which are calculable.
For the more phenomenologically interesting case of $dF_2/d\ln Q^2$
we find from a straightforward calculation
\bea
h_{dg}(\gamma)&=& \frac{4^{\gamma}
\Gamma(1+\gamma)}{(1-\gamma)\Gamma(2-\gamma)},\\
h_{2d}(\gamma) &=& \frac{(1+3/2\gamma-3/2\gamma^2)}{1-\gamma}
\frac{4^{-\gamma}
\Gamma^4(2-\gamma)\Gamma^2(1+\gamma)}{\Gamma(4-2\gamma)\Gamma(2+2\gamma)},\\   
h_{2g}(\gamma)&=& \frac{3/2(2-3\gamma+3\gamma^2)}{3-2\gamma}\frac{
\Gamma^3(1+\gamma)\Gamma^3(1-\gamma)}{\Gamma(2+2\gamma)\Gamma(2-2\gamma)},
\label{eq:hdefs}
\eea
and the equivalence in Eq.(\ref{eq:LOequiv}) is easily verified. 

The implications of this turn out to be rather interesting. Each of these
effective coefficient functions can be expanded as a power series in $\gamma$
about $\gamma=0$, where each has been normalized so that $h(0)=1$.  
For $dF_2/d\ln Q^2$ we obtain, 
\beq
h_2(\gamma(\bar\alpha_S/N)) = 1+ 2.17\gamma+2.30\gamma^2+5.07\gamma^3+
3.58\gamma^4+8.00\gamma^5+\cdots \label{eq:h2def}
\eeq
In order to interpret this we need to know more about $\gamma$. 
Strictly speaking, these expressions are all derived within the LO in $\ln(1/x)$
framework. In this case the gluon anomalous dimension is given by the 
LO BFKL equation. This results in the power-series expansion  
\beq
\gamma(\alpha_S/N) = 
\frac{\bar \alpha_S}{N} + 2.4 (\frac{\bar \alpha_S}{N})^4 + 2
(\frac{\bar \alpha_S}{N})^6 + 17(\frac{\bar \alpha_S}{N})^7+\cdots,
\eeq
where $\bar \alpha_S =(3/\pi)\alpha_S$. In $x$ space this results in a 
splitting function
\beq
xP_{gg}(\alpha_S,x) = 
\bar \alpha_S + 2.4 (\frac{\bar \alpha^4_S \ln^3(1/x)}{3!}) + 2
(\frac{\bar \alpha_S^6\ln^5(1/x)}{5!}) + 
17(\frac{\bar \alpha^7_S\ln^6(1/x)}{6!})+\cdots.
\eeq
This steep growth of $xP_{gg}$ as $x$ decreases leads to a quickly increasing
small-$x$ gluon distribution as $Q^2$ increases. However, substituting 
$\gamma(\alpha_S/N)$ into Eq.(\ref{eq:h2def}) we see that the effective 
coefficient function also grows quickly at small $N$, or equivalently at small
$x$, and consequently $d\,F_2(x,Q^2)/d\ln Q^2$ grows quite a lot more quickly
than $g(x,Q^2)$ with decreasing $x$.

Expanding the other expressions in Eq.(\ref{eq:hdefs}) in powers of $\gamma$
we obtain
\bea
h_{dg}(\gamma(\bar\alpha_S/N)) &=& 1+ 2.23\gamma+3.49\gamma^2+3.95\gamma^3+
4.22\gamma^4+4.06\gamma^5+\cdots\\
h_{2d}(\gamma(\bar\alpha_S/N)) &=& 1- 0.07\gamma-1.05\gamma^2+3.77\gamma^3-
4.94\gamma^4+6.53\gamma^5-\cdots.
\eea
Hence, $h_{dg}(\gamma)$ has a power series expansion in which all the 
coefficients are positive, and of rather similar size to those in the expansion
of $h_2(\gamma)$, whereas $h_{2d}(\gamma)$ has a series expansion where
the first two terms are small and negative, and higher terms oscillate. 
This has an obvious consequence. To a reasonable approximation all the 
small-$x$ enhancement of $dF_2/d\ln Q^2$ relative to $g(x,Q^2)$ is
generated by the dipole-gluon cross-section, which is therefore 
itself steep relative to the gluon. In the final conversion from the dipole 
cross-section to $dF_2/d\ln Q^2$ there is little change in $x$ dependence.   

\begin{figure}
\centerline{\epsfxsize=0.75\textwidth\epsfbox{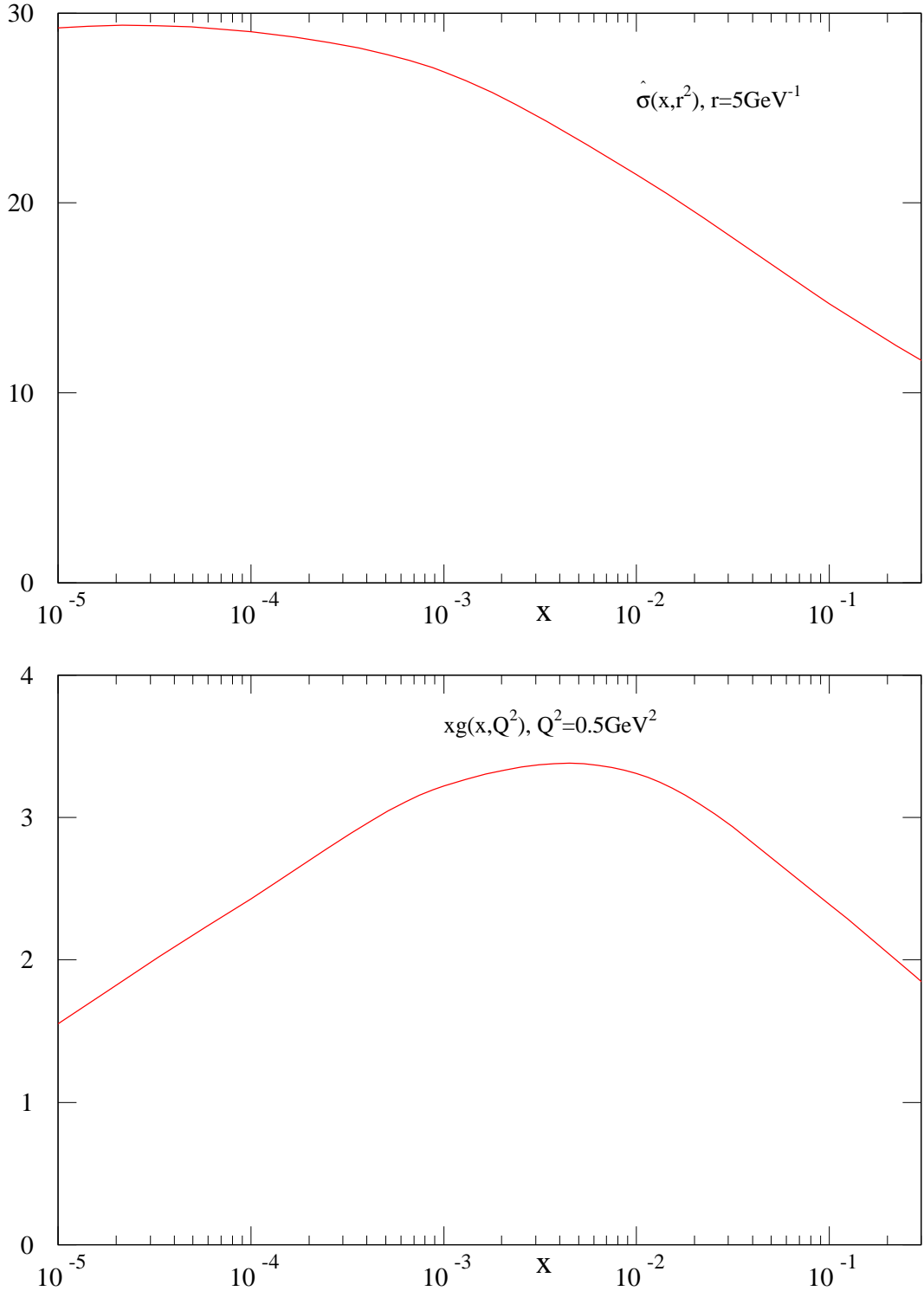}}
\vspace{-3cm}
\caption{$\hat \sigma(x,r^2)$ and $xg(x,Q^2)$ obtained from the original
Golec-Biernat W\"usthoff dipole model \cite{GBW1}. $xf_g(x,k^2)$ has the same 
type of shape as $xg(x,Q^2)$. A flattening dipole at small $x$ 
requires a valence-like integrated or unintegrated gluon distribution.}
\label{fig1}
\end{figure}

In practice the LO BFKL prediction for the gluon is not really used 
in any realistic dipole model fits to data. It predicts a power-like 
behaviour at small $x$ of $x^{\lambda}$, where $\lambda = 4\ln(2) \bar 
\alpha_S = 2.65 \alpha_S$, whereas the sort of power-like behaviour used is
$\lambda \sim 0.25 -0.3$. This means an effective value of $\alpha_S=0.1$
which is extremely low for scales of a few GeV, where $\alpha_S=0.2-0.3$, 
which is where the data 
exist. Hence some models use behaviour implied by higher order corrections 
within the BFKL framework (see e.g. the calculation in \cite{Triant}), 
and others just use models where the gluon is 
more closely associated with standard LO or NLO in $\alpha_S$ perturbative 
QCD. Some are more {\it ad hoc}. This departure from the LO
$k_T$-factorization is already a deviation from the theoretical framework 
within
which the simple dipole picture is valid, but I will essentially disregard 
this 
particular issue throughout this paper. Whatever the inspiration for the 
(unintegrated) gluon distribution, in order to match even the most qualitative
trends of the data one needs a gluon distribution that increases quickly
with $Q^2$ at small $x$, and every model ultimately has an effective
$\gamma(\alpha_S,N)$ that grows quickly at small $N$ and equivalently at 
small $x$. Therefore, the general conclusion on the role of the effective 
coefficient functions made above is always true. Since $\gamma$ is positive
and increasing at small $x$, $dF_2/d\ln Q^2$ is always steeper in $x$
than $g(x,Q^2)$, and this relative steepness is always associated with 
the step taking one from the gluon to the dipole cross-section, with the step
taking one from the dipole cross-section to the physical structure function 
being largely unimportant as far as the shape is concerned.

This conclusion does not seem to have been made previously, but it is easy to 
demonstrate using the very simple dipole model proposed in \cite{GBW1}. 
In this case the dipole cross-section is given by
\beq
\hat \sigma (x,r^2) = \sigma_0(1-\exp(r^2/4(x_0/x)^{\lambda})),
\eeq
where, for the case of the realistic fit which includes charm as a parton 
rather than simply using three light flavours, the parameters obtained from 
the best fit were $\sigma_0=29.2mb, x_0=4\times 10^{-5}, \lambda=0.28$.
This cross-section clearly saturates at large enough $r$ or small enough $x$. 
Using the relationship between the dipole cross-section and the unintegrated 
gluon distribution it is straightforward to obtain
\beq
f_g(x,k^2)=\frac{3\sigma_0}{4\pi^2\alpha_S}k^4(x/x_0
)^{\lambda} e^{-k^2(x/x_0)^{\lambda}},
\eeq
and this expression is indeed used in \cite{GBW2}. Thus, whilst $\hat 
\sigma(x,r^2) \to \sigma_0$ as $x \to 0$, $f_g(x,k^2) \to 0$ as $x \to 0$,
i.e. it has a valencelike behaviour $f_g(x,k^2) \sim x^{0.28}$ as $x \to 0$.
Using the leading twist relationship between the unintegrated and integrated 
gluon distribution, 
$g(x,Q^2)=\int_0^{Q^2}\frac{d k^2}{k^2} f(x,k^2)$, and using fixed 
coupling (the general result does not depend on this), we obtain 
\beq
xg(x,Q^2)=\frac{3\sigma_0}{4\pi^2\alpha_S}\biggl[-Q^2 
e^{-Q^2(x/x_0)^{\lambda}} +(x_0/x)^{\lambda}
(1-e^{-Q^2(x/x_0)^{\lambda}})\biggr],
\eeq
which also behaves like $x^{0.28}$ as $x \to 0$.
This is slightly reminiscent of the valencelike gluons obtained in 
global fits, except that the valence-like behaviour will always set in 
at low enough $x$ in this model (though extremely low $x$ for high $Q^2$),
whereas in the DGLAP approach the valencelike behaviour soon disappears 
with evolution to higher $Q^2$. The behaviour of the dipole cross-section
at large $r$ is compared to that of $g(x,Q^2)$ at low $Q^2$ in Fig. 1, and 
one clearly sees that the eventual flattening of $\hat \sigma(x,r^2)$ 
at low $x$ is accompanied by a distinct turnover in $g(x,Q^2)$, with the 
maximum as high as $x\sim 0.005$ at $Q^2=0.5\GeV^2$. 

It is certainly reasonable to argue that the simple relationship between the 
integrated and unintegrated gluons is not meant to be used in this case,
since higher twist corrections to the gluon will be important. However, 
again this takes us beyond the regime where the strict equivalence between
the dipole picture and the rigorously defined $k_T$-factorization theorem
is valid. Also, even if one doubts the result presented in terms of 
$g(x,Q^2)$, it is certainly the case that $f_g(x,k^2)$ is valencelike.
Hence, this simple example shows that the dipole cross-section becoming
large at a small value of $x$ does not necessarily mean that the gluon
at this value of $x$ is also large. This perhaps clouds the issue of what 
saturation actually means, i.e. does it have to mean a large parton density.
However, it might also go some way towards explaining why fits including
saturation corrections seem to be successful, while standard DGLAP fits 
produce very small (or negative) gluons, and small $x$ and $Q^2$.

In order to provide a definitive answer to this apparent contradiction it
is necessary to undertake some rather more precise work. Although the original 
Golec-Biernat W\"usthoff dipole model was successful with the HERA data 
available at the time, it now produces a qualitatively good fit at best.
And some of the more recent fits are direct attempts to improve it, e.g.
\cite{GBK}. Thus in order to make quantitative conclusions it is necessary to 
relate the gluon distribution and dipole cross-section from a genuinely good 
fit to current data.

\section{The Model for the Gluon Distribution}

In order to make a detailed investigation I will work on the principle that 
reasonably high $Q^2$ the gluon distribution will behave exactly 
according to standard fixed order DGLAP evolution. Hence, I propose a 
simplified model for the gluon which accurately
represents this but contains a minimum of parameters. 
I will then use the exact relationship between the 
unintegrated gluon distribution and the dipole cross-section in 
Eq.(\ref{eq:dipsection}), along with the standard identity, 
$f_g(x,k^2)= \biggl(\frac{d\,g(x,Q^2)}{d \ln Q^2}\biggr)_{Q^2=k^2}$, 
valid at leading twist, in order to obtain the correct expression for 
$\hat \sigma(x,r^2)$.    
In the small $r^2$ limit $\hat \sigma(x,r^2)$ may be written as
\beq
\hat \sigma(x,r^2) = \frac{4\pi^2}{3}r^2 
\int \frac{d k^2}{k^2} \alpha_S f(x,\mu^2) \sim \frac{\pi^2 \alpha_S}{3}
r^2 g(x,\mu^2), 
\eeq
where $\mu^2 = A/r^2$ \cite{rlimit}. The constant $A$ depends on the precise 
behaviour of the gluon, but  it is always the case that $A\approx 10$.
It should not be used as a free parameter in a fit. This expression is 
sometimes used to relate the gluon distribution and dipole cross-section.
However,
it is only ever approximate and only reasonable for very small $r$.
The completely correct Eq.(\ref{eq:dipsection}) should really be used if
one is going into the regime of large $r$ and or small $Q^2$ and $k^2$. 

In order to attempt to obtain a matching between the dipole model gluon 
distribution and a DGLAP one at large $Q^2$
I define a gluon distribution which behaves like a conventional global fit 
gluon at high scales. I note that to a very good 
approximation at LO in $\alpha_S$ the gluon anomalous dimension is 
$\gamma_{gg}(\alpha_s(Q^2),N) =\bar \alpha_s(Q^2)(1/N-1)$, only differing 
from this approximate form significantly at very high $N$. Hence, 
the LO evolution is given by this anomalous dimension to a good 
accuracy except for fine details at the highest $x$. Furthermore, apparently
by accident the NLO correction to gluon evolution is very small, 
the coefficient of  a possible term of $1/N^2$ happening to be zero. 
It was shown in \cite{DAS} that for a flat input at scale $Q_0^2$ the solution
to the evolution equation for $Q^2>Q_0^2$ using this anomalous dimension 
is roughly 
\beq
xg(x,Q^2)\propto I_0\biggl(\!\biggl(2.4\zeta_0
\log\biggl(\frac{\log((Q^2)/\Lambda_{QCD}^2)}
{\log((Q_0^2)/\Lambda_{QCD}^2)}\biggr)\!\biggr)^{0.5}\biggr)
\exp\biggl(\!-1.5\log\biggl(\frac{\log((Q^2)/\Lambda_{QCD}^2)}
{\log((Q_0^2)/\Lambda_{QCD}^2)}\biggr)\!\biggr),
\label{eq:DAS}
\eeq
where $\zeta_0 = \ln(x_0/x)$, and strictly speaking $x_0$ is the value of
$x$ above which the flat input gluon distribution falls away to zero, 
i.e. $g(x,Q_0^2) \propto \Theta(x_0-x)$. 

This is a very good starting point for a more realistic gluon distribution.
The main modification to be made is to round off the high-$x$ behaviour 
to something more like $(1-x)^5$ rather than a $\Theta$ function at $x_0$, and 
to take account of this in the evolution. Also, I want a gluon that can be 
used all the way down to $Q^2=0$ rather than stopping at some input scale, 
and which tends smoothly to a flat behaviour in $x$ as $Q^2 \to 0$.
This is achieved by modifying Eq.(\ref{eq:DAS}) to 
\bea
xg(x,Q^2)&=& A \biggl(\frac{5}{5+\zeta}\biggr)^2\exp
\biggl(-1.5\log\biggl(\frac{\log((Q^2+Q_0^2)/\Lambda_{QCD}^2)}
{\log((Q_0^2)/\Lambda_{QCD}^2)}\biggr)\biggr)\\
& & \hspace{-1cm}\times\biggl(I_0\biggl(2.4
\biggl(\frac{\eta^2(1-\exp(-\eta/4))}{\eta+2.3}(1-\exp(-\eta))^4
\log\biggl(\frac{\log((Q^2+Q_0^2)/\Lambda_{QCD}^2)}
{\log((Q_0^2)/\Lambda_{QCD}^2)}\biggr)\biggr)^{0.5}\biggr)-1\biggr).\nonumber
\eea
$A \biggl(\frac{5}{5+\eta}\biggr)^2$ is the {\it input}, with $A$ the
normalization. This is simply an empirical modification of an original formula
which was theoretically correct in a slightly idealized framework; and
for moderate and high $Q^2$, i.e. above a few $GeV^2$, the gluon does behave 
very similarly indeed to the standard DGLAP gluons coming from global fits.  
$\Lambda_{QCD}=0.12\GeV$, which for the one-loop coupling gives
$\alpha_S(M_Z^2) = 0.118$, i.e. roughly the correct value, and hence it 
gets the speed of evolution correct.  $Q_0^2 = 0.5 \GeV^2$, and marks
the transition scale around which perturbative evolution is beginning to
break down. $Q_0^2$ and the input shape $(5/(5+\eta))^2$ are chosen to give 
roughly the correct phenomenological shape in $x$ and $Q^2$ for the 
lowish $Q^2$ gluon, but 
neither is at all fine-tuned. $Q_0^2$ takes on a perfectly typical value 
for the scale of nonperturbative physics. It clearly serves the function of 
slowing the evolution $Q^2\sim Q_0^2$, but does this in an 
$x$-independent way. I only make the change $Q^2 \to Q^2 +Q_0^2$, independent
of any consideration of $x$. This model of the gluon is not therefore 
inspired at all by the idea of slowing evolution associated with high parton 
densities at small $x$, and does not contain saturation effects.  

This expression for the gluon is converted into a dipole
cross-section using 
\beq
\hat \sigma(x,r^2) = \frac{4\pi^2}{3} 
\int\frac{d k^2}{k^4} \alpha_S(k^2) f(x,k^2)(1-J_0(kr)),
\eeq
where for consistency $\alpha_S(\mu^2)$
is also slowed at low scales with the same regularization as the gluon. 
\beq
\alpha_S(\mu^2) = \frac{4\pi}{\beta_0 \log((\mu^2+Q_0^2)/\Lambda_{QCD}^2)}.
\eeq
The results of this paper are largely insensitive to the details of how the 
low scale coupling is regularized. 
The resulting dipole cross-section is then put into a fit to data. 
The normalization $A$ is the only really free parameter associated with 
the gluon in this fit.  
\begin{figure}
\begin{center}
\mbox{\hspace{-0.5cm}\epsfig{figure=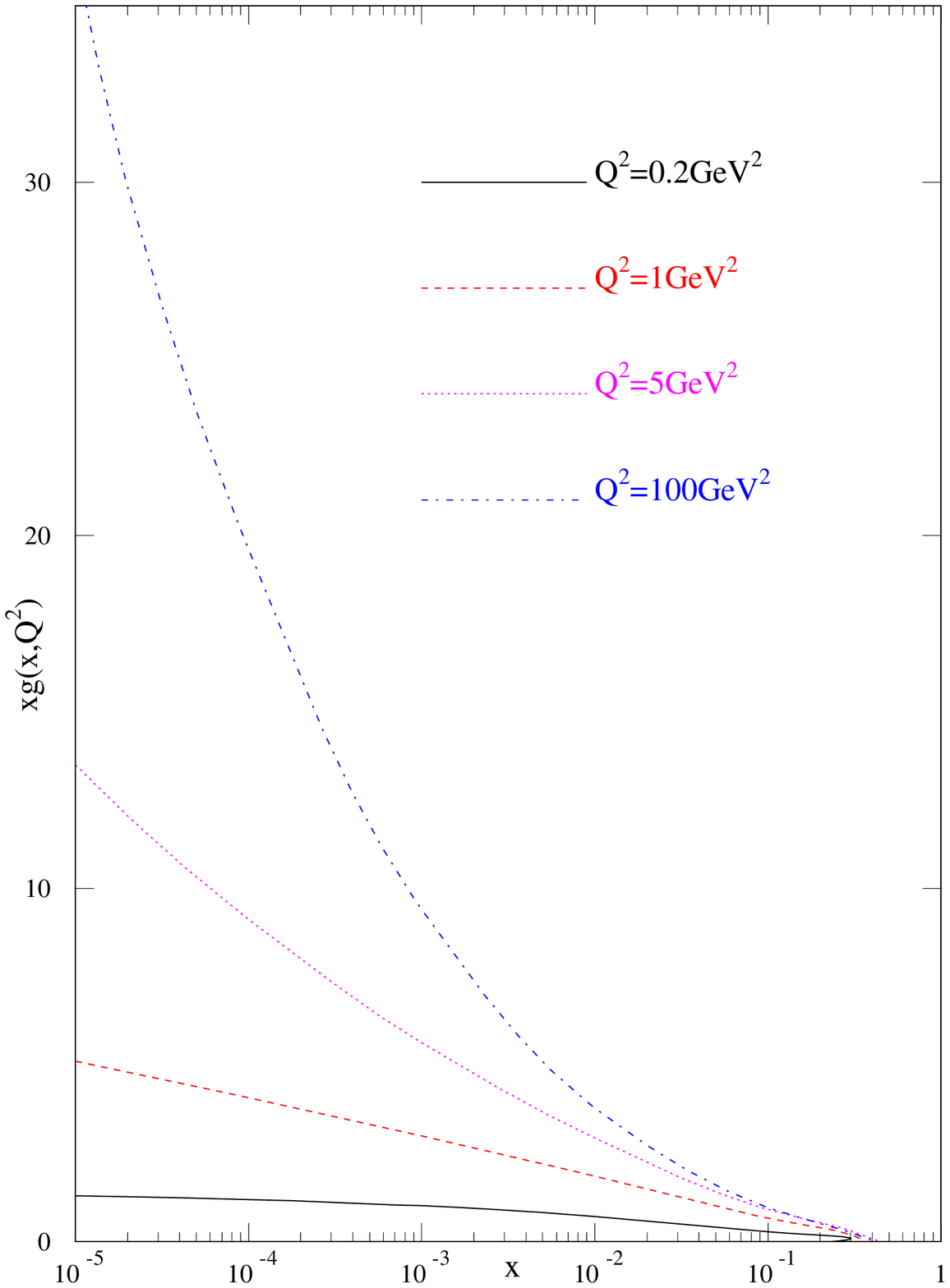,width=8cm}
\epsfig{figure=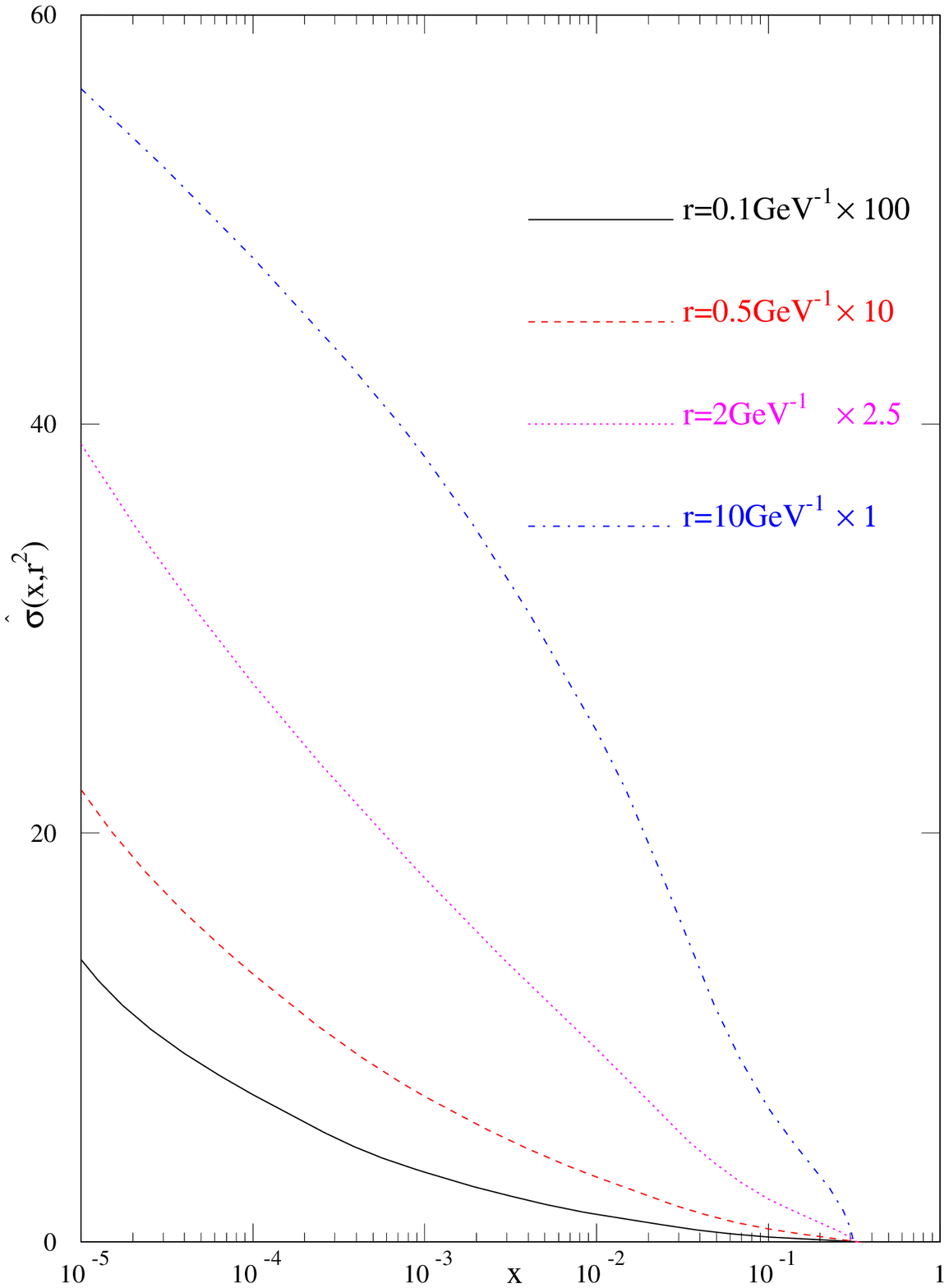,width=8cm}}
\vspace{0.5cm}
\caption{\label{fig2} Comparison of $xg(x,Q^2)$ in my model at various 
$Q^2$ with the corresponding $\hat\sigma(x,r^2)$ at various $r$. 
The normalization is determined by the best fit to data.}
\end{center}
\end{figure}

Since the shape of the gluon has now been determined, we can investigate 
the relative shapes of the gluon distribution and the dipole cross-section
without having to know the value of $A$. This is illustrated in Fig. 2, 
where we 
see $xg(x,Q^2)$ and $\hat \sigma(x,r^2)$ for a variety of values of $Q^2$ 
and $r$. For the largest value of $Q^2$ and correspondingly the smallest 
value of $r$ both $g(x,Q^2)$ and $\hat \sigma(x,r^2)$ rise steeply at small
$x$, and it is difficult to see any particular difference in shape.
However, at the two lower values of $Q^2$, $1\GeV^2$ and $0.2\GeV^2$, 
$g(x,Q^2)$ is clearly flattening out, and is rising very slowly indeed 
in the latter case. For the two larger values of $r$, $2\GeV^{-1}$ and
$10\GeV^{-1}$ (which would naively correspond to $Q^2$ of about $2.5\GeV^2$ and
$0.1\GeV^2$ respectively), 
$\hat \sigma(x,r^2)$ is clearly still rising steeply, and it is
in this regime, which is where any saturation effects are supposed to be 
large, that the influence of the effective dipole-gluon coefficient function 
is clearly seen. It would seem easy to believe that saturation were occuring 
due to a large dipole cross-section, but seems less plausible that this could 
be interpreted as being due to a steeply growing density of gluons.      

\section{Details of the Fit}

There are a number of modifications required compared to the standard 
approach to 
fits to data made within the dipole picture in order to get a truly 
quantitative comparison to the conventional DGLAP approach.
One extremely important issue is the treatment of heavy quarks,
i.e. charm and bottom. These are often ignored in dipole fits. This is 
certainly excusable for the bottom contribution, which 
only really turns on above $Q^2=m_b^2\approx 20 \GeV^2$, and carries a charge 
weighting of 1/9. However, charm contributes about $40\%$ of   
$dF_2/d\ln Q^2$ for $Q^2 > m_c^2$, i.e. it turns on from zero to a very 
sizeable contribution indeed somewhere in the middle of the dipole regime, and
to ignore it is ludicrous. 

However, the charm contribution to inclusive
$F_2(x,Q^2)$ is often left out in dipole model fits. In the original
\cite{GBW1} fits, two fits were made, with and without charm. The latter
actually gave a worse fit, and the form of the dipole cross-section changed,
the saturation parameter $x_0$, i.e the value of $x$ at which saturation 
becomes very important at $Q^2=1\GeV^2$, changed from $x_0=3\times 10^{-4}$
without charm to $x_0=4\times 10^{-5}$ with charm, and the overall magnitude 
of the dipole cross-section decreased significantly, to about $60\%$
the previous value except at very large $r$ or extremely small $x$.     
Neither of these results is surprising. If one is missing up to
$40\%$ of $dF_2/d\ln Q^2$ one would expect an enhancement of
$\hat \sigma(x,r^2)$ and $g(x,Q^2)$ of up to $1.67$.
Since the leading saturation corrections are $\propto g^2(x,Q^2)$, one 
would then expect the saturation effects to be much exaggerated when charm
is absent, as seen.\footnote{Heavy flavours are also absent in the fit in 
\cite{Iancu}. If charm is included then the fit quality does improve slightly,
but again the parameter $x_0$ decreases, from $\sim 4\times 10^{-5}$ to
$\sim 10^{-5}$ in this case \cite{Iancucom}.} For some reason the results of
the dipole fits are habitually cited using the parameter values 
from the fits with charm absent. These parameters are simply wrong, and 
should not be used. Moreover, they clearly suggest that saturation effects are
quite a lot larger than the results obtained from the more correct dipole
fits.   

In order to investigate the importance of the charm contribution to 
inclusive $F_2(x,Q^2)$ I tried performing global fits with this contribution 
set to zero, i.e. mimicking what is done in many dipole fits. 
The procedure for the fit 
was exactly as in the usual MRST fits other than this one modification.
It seems obvious that, in order to counter the absence of a large part of 
the theoretical contribution, the gluon must get bigger at small $x$ to
increase evolution. However, the gluon cannot simply get bigger everywhere
because of the momentum sum rule, so it seemed very likely that $\alpha_S$ 
would also have to get bigger to also try to speed up evolution. 
The results were broadly in line with these expectations, but were rather 
dramatic in other senses.  The main point to note is that the quality of 
the global fit performed in this manner is terrible, with $\chi^2=4000$ 
for $2000$ points, twice that of the normal global fit. At small $x$ it is
impossible to get $dF_2/d\ln Q^2$ consistently correct at all $x$ and $Q^2$.
At low $Q^2$ the gluon wishes to be not too much bigger than normal,
charm not yet being so important in the evolution. However, such a low 
$Q^2$ gluon is then much too small to get evolution correct at higher $Q^2$.
Conversely a gluon large enough for the higher $Q^2$ data is far too big
for the evolution at low $Q^2$. There is no way around this within the 
factorization theorem. Also the increased $\alpha_S$ needed to help the 
small $x$ fit makes the fit to the rest of the data much worse. The quality 
of the fit breaks down everywhere. 

This slightly surprising result may be viewed as a very positive one for 
collinear factorization. It shows that NLO and NNLO DGLAP calculations 
are good enough and constraining enough to determine that charm has 
to be there, and to constrain its mass quite accurately, even without 
using any data directly sensitive to charm. This suggests one should be 
suspicious of 
good qualitative results obtained from calculation where heavy quarks are 
ignored, e.g. the proposed geometric scaling \cite{Stasto}. Such results 
ought to be incorrect, by up to $40\%$, until the heavy flavours are included. 
It is also an indication of the lack of constraints on a theory if the
free parameters can be readjusted to account for such a large change in
the theoretical prescription. Although the input partons in the DGLAP 
approach have a large number of free parameters (it would be very much fewer 
if only small $x$ data were fit), there is only freedom at a given $Q_0^2$.
How one evolves to other $Q^2$ is precisely defined, and this provides a very
strong constraint, as the above discussion illustrates. Even though many 
of the dipole models have few free parameters, they are such that the whole
shape in $x$ and $r^2$, or equivalently $x$ and $Q^2$, can be changed, and
in practice this allows much more freedom.       

Given the above considerations in my dipole fit I include the charm 
contribution, which is 
done by including the wavefunction for the probability for the photon to 
fluctuate into a $c, \bar c$ pair. The only parameter is the charm mass, and   
I use $m_c=1.3\GeV$. I do not include the bottom since it only contributes at
fairly high $Q^2$ and gives a contribution of at most a few percent, which is 
comparable to or less than the errors on the data where it contributes. I note 
that since the inclusion would give a positive contribution, its effect would 
have to be to make the extracted dipole cross-section and resulting gluon
a little smaller.  
  
One also  has to be careful about the precise details of the light quarks 
in the fit. In total three types of diagram enter into the expression for
$F(x,Q^2)$, shown below. 

\vspace{-1.5cm}

\begin{picture}(600,250)(0,0)
\SetColor{Blue}
\SetScale{0.8}
\Photon(20,230)(80,190){5}{6}
\Text(8,180)[]{$\gamma^{\star}$}
\ArrowLine(80,190)(160,190)
\ArrowLine(80,150)(80,190)
\ArrowLine(160,150)(80,150)
\Gluon(80,150)(80,130){5}{2}
\DashLine(80,130)(80,110){5}
\Gluon(80,110)(80,70){5}{3}
\Gluon(80,90)(160,90){5}{5}
\ArrowLine(20,20)(80,70)
\Text(9,9)[]{P}
\ArrowLine(80,70)(160,70)
\ArrowLine(80,70)(160,55)
\ArrowLine(80,70)(160,40)
\CCirc(80,70){10}{Blue}{Blue}
\Text(28,54)[]{$f_{g}(x,k^2)$}
\Photon(220,230)(280,190){5}{6}
\Text(167,180)[]{$\gamma^{\star}$}
\ArrowLine(280,190)(360,190)
\ArrowLine(280,150)(280,190)
\ArrowLine(360,150)(280,150)
\Gluon(280,150)(280,130){5}{2}
\DashLine(280,130)(280,110){5}
\Gluon(280,110)(280,90){5}{2}
\ArrowLine(280,90)(360,90)
\ArrowLine(280,70)(280,90)
\ArrowLine(220,20)(280,70)
\Text(168,9)[]{P}
\ArrowLine(280,70)(360,70)
\ArrowLine(280,70)(360,55)
\ArrowLine(280,70)(360,40)
\CCirc(280,70){10}{Blue}{Blue}
\Text(188,54)[]{$f_{q}(x,k^2)$}
\Photon(420,230)(480,190){5}{6}
\Text(327,180)[]{$\gamma^{\star}$}
\ArrowLine(480,190)(560,190)
\ArrowLine(480,70)(480,190)
\ArrowLine(480,70)(560,70)
\ArrowLine(480,70)(560,55)
\ArrowLine(420,20)(480,70)
\ArrowLine(480,70)(560,40)
\CCirc(480,70){10}{Blue}{Blue}
\Text(350,55)[]{$q(x,Q^2)$}
\Text(328,9)[]{P}
\Text(160,100)[]{$+$}
\Text(320,100)[]{$+$}
\end{picture}

In the dipole picture usually only the left-hand diagram is considered, i.e. 
the whole cross-section comes form the unintegrated gluon within the proton.
However, there is the additional possibility that the unintegrated quark
will emit a gluon which then enters into the same type of scattering process,
as shown in the middle diagram.  In the LO $k_T$-factorization theorem these 
two diagrams contribute to the total as $f_g(x,k^2) + 4/9 f_S(x,k^2)$, 
i.e. it is not just the unintegrated gluon contributing to 
dipole cross-section, but really this combination that should appear in
Eq.(\ref{eq:dipsection}). I shall bear this in mind when investigating the 
results quantitatively. Finally there is the right-hand diagram which
shows the photon scattering from the nonperturbative quark distribution
(gluons could also be radiated off the vertical quark line, but this gives a 
subdominant contribution at small $x$).
Hence, as well as the contribution to the cross section in 
Eq.(\ref{eq:xsectiondipa}) I also include a contribution of the form
$f_{NP} \times Q^2/(Q^2+Q_0^2)$, representing the part of $F_2(x,Q^2)$ coming 
from the right-hand diagram. $f_{NP}$ is a free parameter which in practice 
is small. The final free parameter is the mass of light quarks $m_q$ 
in the expressions for the wavefunctions.  

I perform a fit to H1 \cite{H1}, ZEUS \cite{ZEUS1,ZEUS2} and E665 \cite{E665} 
data for $x<0.01$ and 
$0.5\GeV^2 \leq Q^2 \leq 50\GeV^2$. The last of these is important
since it constrains the $x$-shape of the structure function, and hence 
dipole cross-section, at low $Q^2$ where the HERA data cover only a
relatively narrow range in $x$. It was included in \cite{GBW1}, but has been 
neglected in some more recent fits. I let the data 
normalizations vary within their errors, which is important since the H1 and 
ZEUS data choose to be $\sim 2\%$ different in their normalization. 
The precise range of the data is not that
important, as long as it is fairly wide, since the aim of this paper is not to
provide evidence for my model, or too get as good a fit as possible, but 
to obtain a quantitatively accurate gluon distribution  
from a dipole picture fit. The fit quality would deteriorate outside the two
limits, however, and I will discuss this later.    

The best fit is obtained for $A=10.0$, $f=0.132$ and $m_q^2=0.039\GeV^2$. 
The quality of the fit is $\chi^2 =1.1$ per point. This is comparable to 
the best fits in the previous approaches. It is about as good as one can 
get for the three different data sets, with H1 and ZEUS data 
tending to pull the 
fit in opposing directions. The size of the dipole cross-section
obtained from this fit has already been shown in Fig. 2. One can see that it
exceeds the typical saturation values of $\sim 30mb$ at very small $x$ and 
large $r^2$. However, for comparison I find that with this up-to-date data 
the simple dipole model of
\cite{GBW1} gives $\chi^2=2.5$ per point, and
with best fit parameters of $\sigma_0 = 57.3mb$, $\lambda=0.234$ and 
$x_0=0.00001$.\footnote{Charm has been included in this fit.} The fit using 
my model for the gluon and dipole cross-section 
begins to fail for $Q^2\leq 0.5\GeV^2$,
with the theory overshooting the data, perhaps giving an indication that
some type of saturation corrections could improve matters. The fit also
fails for $Q^2>50\GeV^2$, where the data are mainly at $x>0.001$. This is
again due to the theory overshooting the data, i.e. $d\,F_2(x,Q^2)/d\ln Q^2$
grows too quickly. Saturation is clearly nothing to do with this failure
-- it is a feature of the dipole model with a realistic high-$Q^2$ gluon. I
will address this in greater detail below.   

In order to try to improve my fit, and perhaps push it to lower $Q^2$,
I incorporate one final modification. I include some higher twist corrections 
due to the type of diagrams shown below. 
   
\begin{picture}(600,200)(0,0)
\SetColor{Blue}
\SetScale{0.75}
\Photon(100,150)(180,150){5}{7}
\Photon(420,150)(500,150){5}{7}
\ArrowArc(300,81)(138,30,150)
\ArrowArc(300,219)(138,210,330)
\Gluon(210,40)(210,115){5}{4}
\Gluon(250,40)(250,91){5}{3}
\Gluon(350,40)(350,209){5}{8}
\Gluon(390,40)(390,115){5}{4}
\end{picture}

\noindent These are the contributions due to multiple dipole scattering
with the proton. It was shown in \cite{Muellercd} that the result of 
summing such diagrams in the leading $\ln(1/x)$ limit is
\beq
\hat \sigma(x,r^2) = \sigma_0(1
-\exp(-\hat\sigma_{simp}(x,r^2)/\sigma_0)), 
\eeq
where $\hat \sigma_{simp}(x,r^2)$ is the formula for the dipole cross-section
we  have used so far, i.e. Eq.(\ref{eq:dipsection}).
Hence, we have a formula similar to that used in \cite{GBW1}. However,
the exponentiation is due to the multiple dipole scattering, while
the relationship between a single dipole scattering and the gluon distribution
is unchanged. Hence, it seems that this may be interpreted as dipole
saturation, but not gluon saturation. 

The best fit is now obtained with the
parameters $\sigma_0=146.3mb$, $A=10.1$, $f=0.118$ and $m_q^2=0.0249\GeV^2$.
The large value of $\sigma_0$ and the extremely small change in $A$ make it
clear that the saturation effects are not chosen to be at all significant
by the best fit. The quality of the fit only improves very slightly, as is
obvious since the dipole cross-section 
itself hardly changes. The extrapolation into the 
region $Q^2<0.5\GeV^2$ is not really improved. The fit quality remains 
much the same as long as $\sigma_0 \geq 60 mb$ 
(a proton radius $R_p=1fm$ corresponds to $\sigma_0=60 mb$), 
with $A$ varying by $< 3\%$. For this lower value of $\sigma_0$ the  
extrapolation for $Q^2<0.5\GeV^2$ remains poor. However, it may be argued that 
for this low $Q^2$ no perturbatively inspired model is really correct, and 
nonperturbative physics is essential to describe the data correctly. 

\begin{figure}
\vspace{-1cm}
\begin{center}
\centerline{\epsfxsize=0.75\textwidth\epsfbox{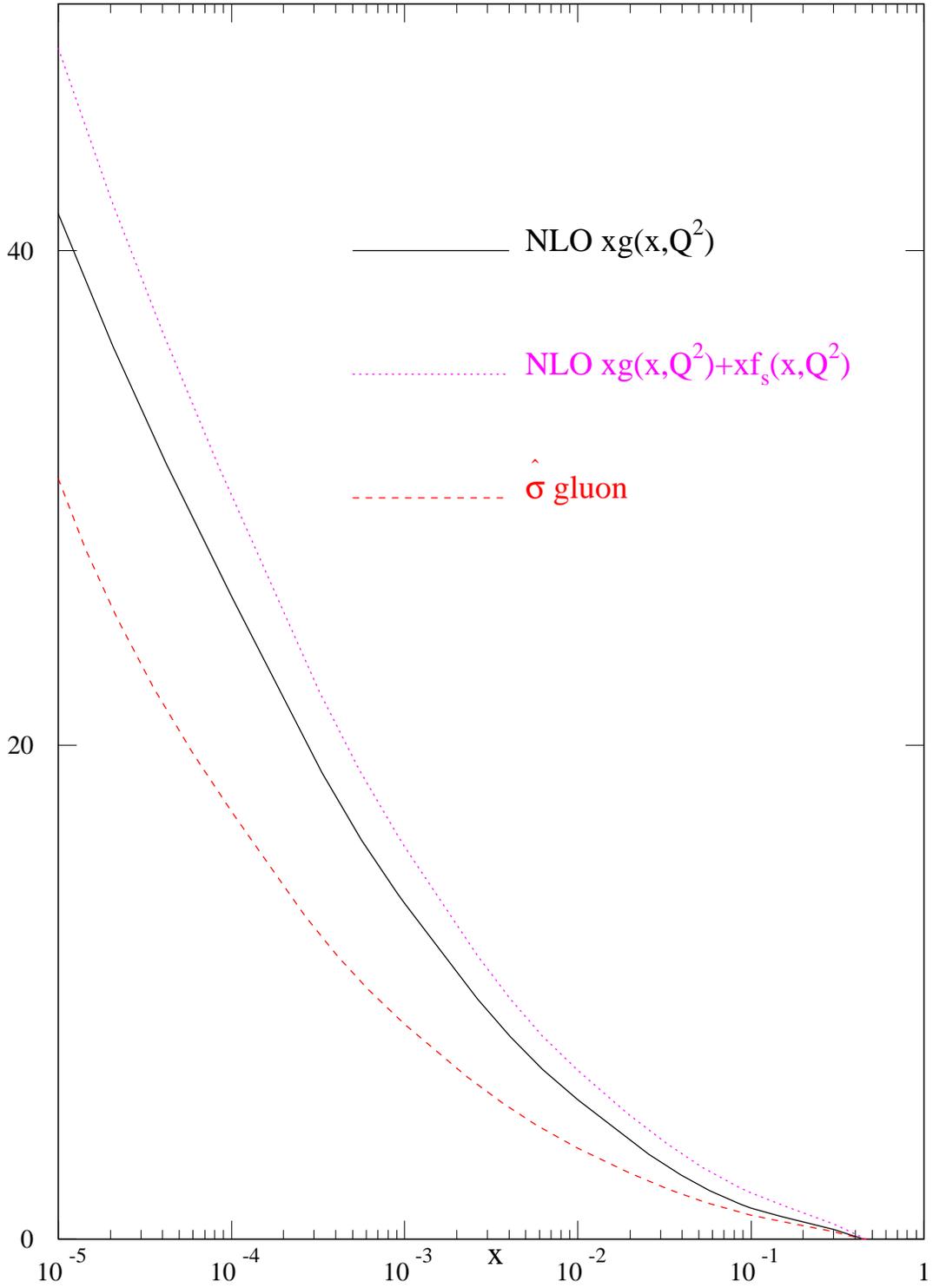}}
\vspace{-3cm}
\caption{\label{fig3} Comparison of the gluon distribution obtained from the 
dipole model fit to the gluon distribution and to $xg(x,Q^2)+ 4/9xf_s(x,Q^2)$
obtained from a conventional NLO global fit. All partons are shown for 
$Q^2=50\GeV^2$.}
\end{center}
\vspace{-1cm}
\end{figure}

Now that we have the parameter $A$ describing the normalization of the 
input gluon distribution we can compare to conventional gluons. 
The value of $A= 10.3$ leads to the gluon distribution shown in Fig. 3
for $Q^2=50\GeV^2$. This is a value of $Q^2$ where the data are still
being fit using the dipole approach, but where saturation effects should
have become negligible except perhaps at the very lowest $x$. 
Hence, this gluon should really be very similar to a conventional gluon 
obtained from the factorization theorem at this $Q^2$.   
It is compared to the MRST2002 NLO gluon \cite{MRSTerror2} in order to test 
this equality. Clearly it fails quite badly, being approximately  
$0.65-0.75$ of the DGLAP gluon.\footnote{Also the MRST NLO gluon is  
relatively small compared to some other NLO gluons.} 
This is even more striking when one remembers
that it should really be compared to $g(x,Q^2)+4/9f_S(x,Q^2)$. In this case 
the factor is now $0.5-0.65$. The biggest suppression in proportional terms
is at high $x$.   

Hence, the gluon obtained from the dipole model fit does not match onto 
the standard DGLAP gluon at high $Q^2$, where they should 
converge. Presumably the DGLAP gluon is correct at $Q^2=50\GeV^2$ since, 
after all, it is producing the correct slope $d\,F_2(x,Q^2)/d\ln Q^2$ to fit
a lot of accurate data at and above this scale  
within what should be a reliable 
theoretical framework. At this sort of scale neither saturation corrections 
nor resummation corrections in $\ln(1/x)$ should be important until 
very small $x$. 
Hence, the dipole motivated fit, with its gluon 
mismatch of up to $50\%$, is quite considerably inaccurate.  
Examining the two competing gluons at low $Q^2$ we find that the dipole 
fit gluon is much smaller than DGLAP at moderate $x$
but at low $Q^2$ eventually becomes bigger at very small $x$, due to the
fact that my replacement of $Q^2 \to Q^2+Q_0^2$ slows the DGLAP evolution for 
$Q^2$ not too much greater than $Q_0^2$, 
and the evolution leads to the biggest 
absolute changes at small $x$. In order to decide how far we should trust 
the gluon coming from the dipole fit we have to understand the relative 
behaviour of this and the DGLAP gluon. It is difficult to say which should be
more reliable at small $x$ and low $Q^2$ until we understand why they 
fail to match at high $Q^2$.  

It is actually not too difficult to do this. The mismatch comes  
from the effective coefficient functions or splitting functions in the 
dipole approach.
Let us consider $dF_2/d\ln Q^2$. This is controlled by the gluon and the 
anomalous 
dimension  
$\gamma^{DIS}(\alpha_S(Q^2),N)$. For my model for the gluon $\gamma_{gg}
(\alpha_s(Q^2),N) =\bar \alpha_s(Q^2)(1/N-1)$, which is a very good 
approximation to the LO or NLO DGLAP anomalous dimension 
(or NNLO even, at fairly large $Q^2$). 
Substituting into Eq.(\ref{eq:h2def}) we obtain for the dipole motivated fit
\beq
\gamma^{DIS}_{dip}(\alpha_S(Q^2),N) \approx 
\frac{\alpha_S(Q^2)2N_f}{6\pi}
\biggl(1+2.17\bar\alpha_S(Q^2)\biggl(\frac{1}{N}-1\biggr)+2.30
\bar\alpha^2_S(Q^2)
\biggl(\frac{1}{N}-1\biggr)^2 + \cdots\biggl).
\label{eq:h2wrong}
\eeq
At reasonably high $Q^2$, and not too low $x$, the first two terms dominate 
the evolution. However, the expression using the exact result for the 
anomalous dimensions is 
\beq
\gamma^{DIS}_{exact}(\alpha_S(Q^2),N) = \frac{\alpha_S(Q^2)2N_f}{6\pi}
\biggl((1-1.08N + \cdots) +2.17\bar\alpha_S(Q^2)\biggl(\frac{1}{N}-3.05 +
\cdots\biggr)+ \cdots\biggl),
\eeq 
where I have expanded the exact LO and NLO anomalous dimensions about $N=0$.  
More precisely, the exact LO splitting function $\frac{\alpha_S N_f}{2\pi}
(1-2x+x^2)$ is replaced by $\frac{\alpha_S N_f}{3\pi}\delta(1-x)$, while the 
full expression for the NLO splitting function is replaced by 
$\frac{\alpha_S N_f}{3\pi}2.17(1/x-\delta(1-x))$. 
Both of the first two terms are a lot bigger in this approximation in the
dipole approach.
Important corrections which are sub-leading in $\ln(1/x)$ are left out of the 
quark anomalous dimensions and splitting functions, significantly increasing 
the speed of evolution for a given gluon distribution. Alternatively, when
performing a fit using this effective splitting function one obtains     
a smaller gluon than one should, particularly at moderate $x$. 
If one goes to very small $x$ (i.e. smaller $N$), the difference
between the correct and effective anomalous dimension at LO and NLO (and NNLO)
in $\alpha_S$ becomes less significant. Hence, the smaller $x$ gluon can be 
nearer the DGLAP result than the moderate $x$ gluon, and the gluon appears 
steeper than it should be. 

This overestimate of the low order in $\alpha_S$ terms can also maintain this
shape of the gluon at small $x$, even when considering the extra terms in the 
dipole anomalous dimension compared to fixed order DGLAP. 
The terms in the effective splitting
function at higher orders in $\alpha_S$ contain parts of the form  $1/N^m$,
which in $x$-space become $\ln^{m-1}(1/x)$. These give a contribution to 
$d\,F_2(x,Q^2)/d\ln Q^2$ of the form  
\beq
\int^1_{x} \frac{dz}{z}\alpha_S^n \ln^m(1/z)\,g(x/z,Q^2).
\eeq
We see that the form of the convolution means that the 
largest values of the splitting function at small $z$,  
$\ln^m(1/z)$, are coupled with the largest $x$
in $g(x,Q^2)$. However, the overestimate of the low order in $\alpha_S$
splitting functions has led to the high and moderate $x$ gluon 
being much smaller than it should be. This minimizes the effect of the 
extra terms in the dipole splitting function coming from the $\ln(1/x)$ 
resummation and means that, even with these 
higher orders in $\alpha_s^n \ln^m(1/x)$ terms, the small $x$ gluon can be 
steeper than it should be.

\begin{figure}
\vspace{-1.5cm}
\begin{center}
\centerline{\epsfxsize=0.75\textwidth\epsfbox{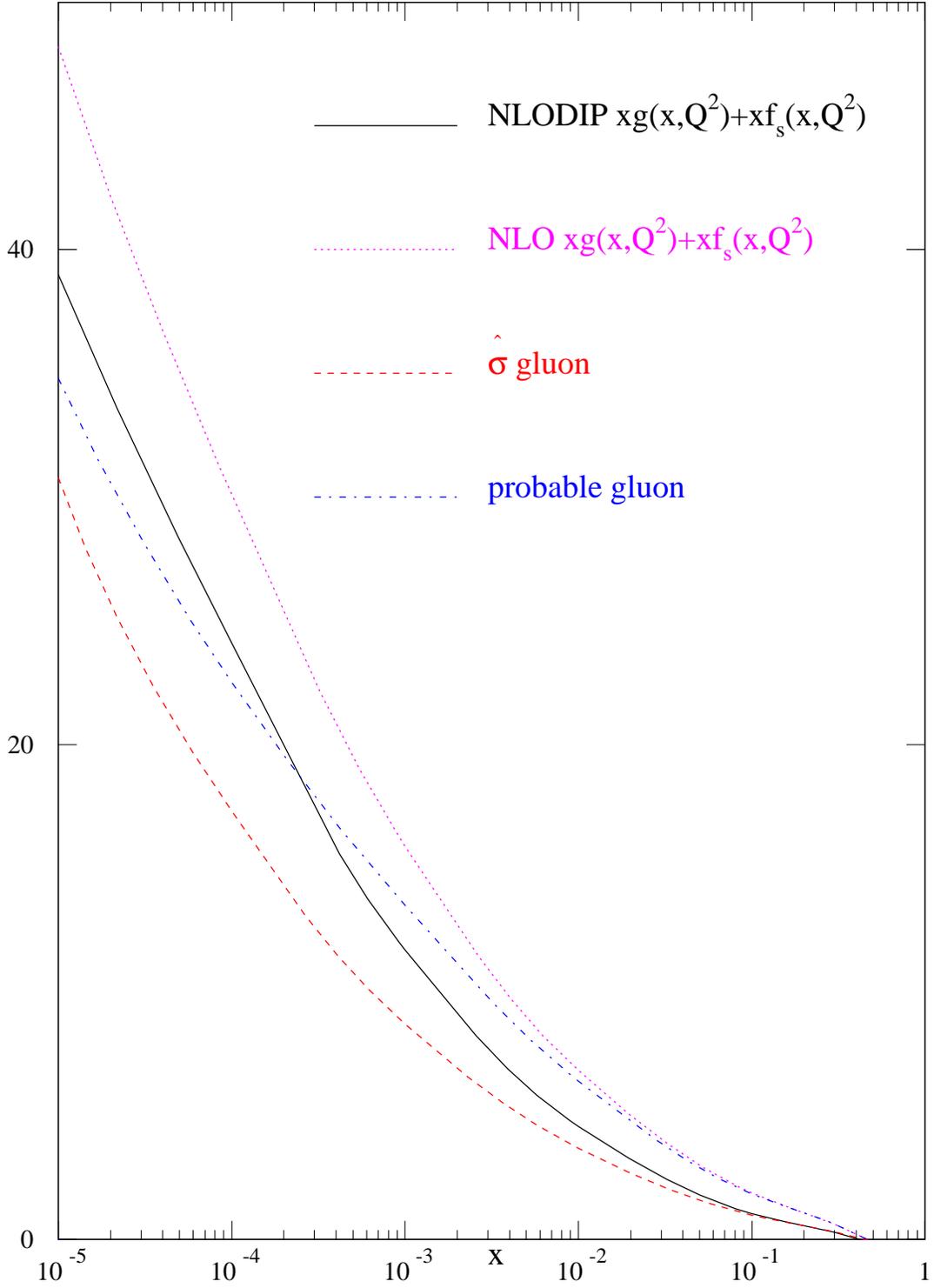}}
\vspace{-3cm}
\caption{\label{fig4} Comparison of the gluon distribution obtained from the 
dipole model fit to $xg(x,Q^2)+ 4/9xf_s(x,Q^2)$
obtained from a conventional NLO global fit
and $xg(x,Q^2)+ 4/9xf_s(x,Q^2)$ from a NLO global fit where
the quark-gluon splitting functions have been used in the same small-$x$
limit as in the dipole approach (NLODIP). 
Also shown is the general probable form of 
a ``correct'' gluon which extrapolates from NLO at high $x$ to something
a little smaller at small $x$. All partons are shown for 
$Q^2=50\GeV^2$.}
\end{center}
\vspace{-1cm}
\end{figure}

So the difference in the dipole and DGLAP 
gluons is largely due to the difference in the effective splitting 
functions. This can be qualitatively verified by directly 
modifying the DGLAP splitting
functions in a conventional global fit, i.e. using the form in 
Eq.(\ref{eq:h2wrong}) for the LO and NLO splitting functions. 
The resulting $xg(x,Q^2)+ 4/9xf_s(x,Q^2)$ for the best fit is shown in Fig. 4.
In this case the distribution is indeed 
much smaller at high and moderate $x$, in fact almost identical to that
obtained from the dipole fit, but becomes steep at low $x$. 
This distribution is exactly what we would expect. It becomes larger
than that in the dipole fit at very small $x$ because this fit is missing 
the contributions to the effective splitting function at 
${\cal O}(\alpha_S^3)$ and beyond. These are positive, speeding the evolution,
and their absence allows the dipole gluon to be a bit smaller at the 
lowest $x$. These 
terms, or at least their full form, should be there in a complete theory, 
so the ``true'' gluon
should be the standard NLO gluon at high and moderate $x$, but a little 
below this at the lowest $x$. A probable, ``correct'' gluon of this form is 
also shown in the figure. Indeed, the NNLO gluon is a little smaller than
the NLO gluon at the smallest $x$, and tends towards the probable gluon. 
This more ``correct'' gluon is hence a rather different shape 
from the dipole gluon, and this difference in shape would increase as $Q^2$ is 
lowered.    

Now that we clearly understand the reason for the mismatch between the 
dipole gluon and the DGLAP gluon at $Q^2=50\GeV^2$
we can also understand why the dipole fit modelled on a gluon distribution 
that behaves correctly fails at $Q^2>50\GeV^2$. In this region the  
contribution at moderate $x$ to $d\,F_2(x,Q^2)/d\ln Q^2$ coming from 
Eq.(\ref{eq:h2wrong}) is just too large when combined with 
a normal gluon evolution. In order to get a good fit to the structure function 
data, the gluon at 
$x\sim 0.01$ must actually fall with $Q^2$ for $Q^2>50\GeV^2$. 
This is simply incorrect phenomenologically, but is achieved 
accidentally in some dipole models. Similarly the  
very good fit at $Q^2 \sim 20-50\GeV^2$ in dipole model fits is 
achieved with the wrong gluon, i.e. one cannot rectify a discrepancy of 
up to $50\%$ over a short evolution range. 
At $Q^2 \sim 1\GeV^2$ the missing terms in the splitting function 
Eq.(\ref{eq:h2wrong}) are still by no means negligible, so the dipole 
extracted gluon cannot be truly accurate here. 
It is also true that in this region there is no good reason to believe that 
the DGLAP gluons are very accurate either, and indeed they look rather odd.
A quantitatively correct gluon in this range would require
a more complete theoretical prescription than either approach 
(and possibly any existing competitor) currently provides.   

\section{Conclusions}

In order to obtain correct results from a fit to structure function data 
one has to be very careful. It is not difficult to obtain a good fit
to the data, which are very smooth in $x$ and $Q^2$, 
and many people have done so since the HERA data began to appear.
It is even possible to do so using physics arguments that are demonstrably 
wrong. It requires far more care to obtain genuinely meaningful results 
with physical interpretations that are in any sense quantitatively correct,
and input quantities that are 
determined to an accuracy where they may reliably be used in 
predicting other quantities. In this paper I have  
performed a fit using the dipole framework, and related this to the standard
leading twist gluon distribution as accurately as possible in order to 
try to understand the seemingly inconsistent results of large, growing
distributions at small $x$ in the dipole approach, and small $x$ 
valence-like and sometimes negative distributions in standard perturbative 
approaches at NLO and NNLO. In doing this I have investigated the degree
of precision with which distributions can be extracted using the dipole 
approach and the amount of faith we should have in the quantitative 
conclusions of such fits.    

One major point to make, which should be self-evident but is 
very commonly ignored, is that when fitting to the inclusive structure 
function data one must use heavy quarks in the theoretical framework for the  
fits. The charm contribution to the structure function 
comprises up to $40\%$ of $d\,F_2(x,Q^2)/d\ln Q^2$ and alters the
qualitative form of results. In fact, the standard DGLAP fit fails completely
if this is left out. 
However, many dipole fits disregard it, overestimating the 
size of the dipole cross-section and the scale at which saturation occurs,
even though the fit is good. Also, it is pointless to show the success of 
the model in predicting the diffractive structure function, if charm
is ignored in both calculations, as is sometimes done; it must be included in 
the extraction of the dipole cross-section and in the calculation of the 
diffractive cross-section (and certainly not in just one of the two, as 
is also sometimes done). If the correct inclusion of charm improves any 
results it might add weight to the evidence for a particular 
theoretical prescription. Conversely, it seems suspicious if the inclusion 
of charm makes results worse. 
 
I discover two reasons which partially explain the apparent discrepancy
between steep dipole cross-sections and valence-like gluons, which are 
nothing to do with any real discrepancy between the DGLAP approach and the 
dipole-motivated approach. Firstly, it is more appropriate to think of the 
dipole cross-section as related to the combination $xg(x,Q^2)+4/9xf_s(x,Q^2)$
rather than just the gluon. This is because unintegrated quarks in the nucleon
can radiate gluons which then go on (possibly with further radiation) to
take part in the scattering with the dipole. In the LO $k_T$-factorization
theorem the gluon and quark contribute to this type of process in the
combination above. This means that, even if at low scales the gluon is 
valence-like, the dipoles can pick up a steep behaviour from the quarks. 
   
Also, the effective coefficient function governing the size of structure 
functions in terms of the gluon in $k_T$-factorization may be unambiguously
split into a structure function-dipole part and a dipole-gluon part. 
The full coefficient function (or splitting function) leads to a 
significant enhancement
of the growth with decreasing $x$, and it is found that essentially all of 
this appears in the dipole-gluon component. Hence, for a given gluon 
anomalous dimension there is a calculable coefficient function which causes 
the dipole cross-section to be considerably steeper than the gluon 
distribution. 

Both these effects are in the direction needed in order to 
reconcile the DGLAP approach and the dipole approach. However, in order 
to test 
fully their compatibility I have constructed a model for the gluon distribution
which evolves quantitatively like a DGLAP gluon for 
$Q^2 \gg 0.5\GeV^2$, but where
the evolution slows down at low $Q^2$ so that for $Q^2<0.5\GeV^2$
$xg(x,Q^2) \sim Q^2$ and becomes flat in $x$. This slowing of the evolution 
is achieved only by altering $Q^2$ in the expression, making no special case 
of small $x$ and hence not invoking saturation type arguments. Using such a 
gluon, a very good fit was obtained for $0.5 \GeV^2 <Q^2 <50 \GeV^2$.
Above $Q^2=50\GeV^2$ a gluon with DGLAP
type evolution used within the dipole approach becomes incompatible with data. 
The predicted cross-sections are too big at small $x$ for 
$Q^2 < 0.5\GeV^2$ and $x\sim 10^{-5}$, and some reduction is necessary here,  
possibly a sign of saturation.     
However, the gluon for the good fit is small, and not very steep at low $Q^2$.
For $x> 10^{-5}$ and sensible values of $R_p$ we never have the condition 
$\frac{\alpha_s(Q^2)\pi xg(x,Q^2)}{Q^2R_p^2}\sim 1$, i.e.
the naive saturation requirement \cite{MQ}. 

I obtain the important result that the extracted gluon is 
much too small to match to a genuine DGLAP gluon at high $Q^2$. 
This real discrepancy between the DGLAP approach,
which must be correct to a good accuracy for $Q^2$ above  $50 \GeV^2$ 
(at least until very small $x$), and the dipole approach
can be seen to be due very largely 
to inaccuracies in effective splitting functions or coefficient
functions used when relating the gluon or dipole cross-sections to
structure functions. They are expressions that are only really valid in the 
leading $\ln(1/x)$ limit, and comparison with the exact perturbative 
coefficient functions and splitting functions shows clearly 
that they give structure functions which are too large. 
This affects both the size and the shape of
the gluon and dipole cross-section extracted, and the error is greatest at the 
moderate $x$ values
where the DGLAP gluon is most reliable, rather than at very small $x$. 
The same problems in relating the dipole cross-section to the structure 
function
exist at smaller $Q^2$, so even though the DGLAP gluon certainly becomes
unreliable at low enough $Q^2$, the dipole cross-section and the resulting 
gluon are not truly reliable either.   

Hence, part of the discrepancy between the dipole approach and the 
conventional DGLAP approach is only an apparent discrepancy 
-- the dipole
cross-section being rather steeper than the gluon distribution at small $x$,
though this means that one cannot immediately regard saturation due to a 
large dipole cross-section as being quite the same as saturation
due to a large gluon distribution. However, part of the discrepancy is real,
with the effective coefficient function allowing one to calculate the 
structure function from the dipole cross-section missing very important 
contributions which are present
in the exact order-by-order in $\alpha_S$ calculations. These contributions 
are really there, and should not be ignored. 
This consistent inaccuracy in relating the true dipole cross-section to 
the structure function data means that one cannot have real faith 
in the quantitative size and shape of the extracted dipole 
cross-sections, and can only treat any claims about the suitability of a 
particular theoretical foundation  based on a fit to data as justified in 
very qualitative terms. It has long been realized that one must go 
beyond LO $k_T$-factorization theory in a calculation of the gluon to 
get any sort of reasonable quality of fit, 
but it is necessary to do likewise for the 
wavefunctions in order to be at the level where one has genuinely 
quantitative results. 

There are various possible avenues. Much work has been done on calculating 
the NLO $k_T$-factorization theory impact factors for photon-gluon scattering
\cite{NLOimpact} to go along with the NLO gluon kernel \cite{NLOBFKL}, 
and these would be 
useful in extending the validity of the formalism. The impact factors with
exact gluon kinematics have already been calculated \cite{Bialas1}, and these
could also give useful information about how to improve the calculational
framework. It would be particularly interesting to compare these results with 
the complete NLO impact factors, once they are known,
to see how well they predict the complete 
NLO contribution. If they are successful in doing this one might hope they
would be a fairly accurate representation at even higher orders. However,
I feel that, even if one is only fitting HERA data at lowish $x$, it is vital 
to use some calculational framework which combines both the leading
terms in a $\ln(1/x)$ expansion and the leading terms in an order by order 
in $\alpha_S$ expansion (along the lines of that used in e.g.
\cite{Thorne}) to be truly accurate, since the latter are always 
important even at very small $x$. Certainly, the use of corrections to the 
coefficient functions, such as those calculated using the exact gluon 
kinematics in \cite{Bialas1}, do increase the overall normalization 
of the gluon for a given structure function, as is required to obtain a 
closer match to the DGLAP gluon at high $Q^2$. However, as shown in 
\cite{Bialas}, the simple dipole picture does not really apply beyond LO in 
the $k_T$-factorization theory, due to the lack of diagonalization of the 
cross-section in the transverse position $r$, and such calculations are still 
within the spirit of the $k_T$-factorization theorem, but are more difficult 
to interpret in terms of the dipole picture. Hence, constructing a 
quantitatively accurate dipole picture cross-section seems to be a 
particularly challenging problem.

\section*{Acknowledgements}

I would like to thank the participants of the discussion meetings on 
evidence for saturation at DIS04, XII International Workshop on Deep Inelastic
Scattering, \v Strebk\'e Pleso, Slovakia, April 2004, and at Low-x 2004, 
Prague,
Czech Republic, Sept. 2004  for useful discussions. I would like to thank
the Royal Society for the award of a University Research Fellowship.


\newpage

\begin{thebibliography}{xx}


\bibitem{MRST2001}A.~D. Martin, R.~G. Roberts, W.~J. Stirling and
R.~S. Thorne, Eur. Phys. J. {\bf C23} (2002) 73.

\bibitem{CTEQ6} CTEQ Collaboration: J. Pumplin {\it et~al.}, JHEP 0207:012 
(2002).

\bibitem{MRSTerror2}  A.~D. Martin, R.~G. Roberts, W.~J. Stirling and R.~S. 
Thorne, Eur. Phys. J. {\bf C35} (2004) 325.

\bibitem{BFKL}      
L.~N. Lipatov, Sov. J. Nucl. Phys. {\bf 23} (1976) 338;\\
E.~A. Kuraev, L.~N. Lipatov,  V.~S. Fadin, Sov. Phys. JETP 
{\bf 45} (1977) 199;\\
I.~I. Balitsky,  L.~N. Lipatov, Sov. J. Nucl. Phys. {\bf 28} (1978) 338.

\bibitem{Frankfurtcd}
L.~L. Frankfurt and M.~I. Strikman,
Phys.\ Rept.\  {\bf 160}, (1988) 235.


\bibitem{Muellercd}
A.~H. Mueller,
Nucl.\ Phys.\ {\bf B335}, (1990) 115.

\bibitem{Nikolaevcd1}
N.~N. Nikolaev and B.~G. Zakharov,
Z.\ Phys.\ {\bf C49}, (1991) 607.

\bibitem{Nikolaevcd2}
N.~N. Nikolaev and B.~G. Zakharov,
Phys.\ Lett.\ {\bf B332}, (1994) 184; Z.\ Phys.\ {\bf C64}, (1994) 631.

\bibitem{GBW1}
K. Golec-Biernat and M. Wusthoff,
%
Phys.\ Rev.\ {\bf D59}, (1999) 014017.

\bibitem{GBW2}
K. Golec-Biernat and M. Wusthoff,
Phys.\ Rev.\ {\bf D60}, (1999) 114023.

\bibitem{Forshaw}
J.~R. Forshaw, G. Kerley and G. Shaw, 
Phys.\ Rev.\ {\bf D60}, (1999) 074012.

\bibitem{McDermott}
M. McDermott, L. Frankfurt, V. Guzey and M. Strikman,
Eur.\ Phys.\ J.\ {\bf C16}, (2000) 641.

\bibitem{GBK}
J. Bartels, K. Golec-Biernat and H. Kowalski,
Phys. Rev. {\bf D66}  (2002) 014001.   

\bibitem{GLLM}
E. Gotsman, E. Levin, M. Lublinsky and U. Maor, Eur. Phys. J. {\bf C27}
(2003) 411. 

\bibitem{Iancu}
E. Iancu, K. Itakaru and S. Munier, 
Phys. Lett. {\bf B590}, (2004) 199. 

\bibitem{Forshaw1} 
J.~R. Forshaw, and G. Shaw, JHEP {\bf 12} (2004) 052.   

\bibitem{MQ} 
A.~H. Mueller and J. Qiu, Nucl. Phys. {\bf B268} (1986) 427.

\bibitem{LR}
E.~M. Levin and M.~G. Ryskin, Phys. Rep. {\bf 189} (1990) 267.

\bibitem{BKetc}
I. Balitsky, Nucl. Phys. {\bf B463} (1996) 99;\\
Yu. V. Kovchegov, Phys. Rev. {\bf D60} (1999) 034008;\\
H. Weigert, Nucl. Phys. {\bf A703} (2002) 823;\\
E. Iancu, A. Leonidov and L. McLerran, Nucl. Phys. {\bf A692} (2001) 583;
Phys. Let. {\bf B510} (2001) 133;\\
E. Ferreiro, E. Iancu, A. Leonidov and L. McLerran, Nucl. Phys. {\bf A703}
(2002) 489.  

\bibitem{classdip}
F. Hautmann, Z. Kunszt and D. Soper, Nucl. Phys. {\bf B563} (1999) 153;\\
W. Buchmuller, T. Gehrmann and A. Hebecker,  Nucl. Phys. {\bf B537} 
(1999) 477.  

\bibitem{Bialas} 
A. Bialas, H. Navalet and R. Peschanski, Nucl. Phys. {\bf B593},
(2001) 438.

\bibitem{kttheorem}
S. Catani, M. Ciafaloni and F. Hautmann,
Nucl.\ Phys.\ {\bf B366}, (1991) 135;\\
J.~C. Collins and R.~K. Ellis,
Nucl.\ Phys.\ {\bf B360}, (1991) 3.

\bibitem{CatHaut}
S. Catani and F. Hautmann,
Nucl.\ Phys.\ {\bf B427}, (1994) 475.  

\bibitem{Triant} 
D.~N. Triantafyllopoulos, Nucl. Phys. {\bf B648} (2003) 293.

\bibitem{rlimit}
L. Frankfurt, G.~A. Miller and M. Strikman, Phys. Lett. {\bf B304} (1993) 1;\\
B. Blattel, G. Baym, L. Frankfurt and M. Strikman, Phys. Rev. Lett.
{\bf 71} (1993) 896;\\
L. Frankfurt, A. Radyushkin and M. Strikman, Phys. Rev. {\bf D55} (1997) 98.   

\bibitem{DAS}
R.~D. Ball and S. Forte, Phys. Lett. {\bf B335} (1994) 77. 

\bibitem{Iancucom} Not published anywhere but results reported by E. Iancu
in discussion meetings at DIS04, XII International Workshop on Deep Inelastic
Scattering, \v Strebk\'e Pleso, Slovakia, April 2004, and at Low-x 2004, 
Prague, Czech Republic, Sept. 2004. 

\bibitem{Stasto}
A.~M. Stasto, K. Golec-Biernat and J. Kwiecinski,
Phys.\ Rev.\ Lett.\  {\bf 86}, (2001) 596.

\bibitem{H1} H1 Collaboration: C. Adloff {\it et~al.}, Eur. Phys. J.
{\bf C21} (2001) 33.

\bibitem{ZEUS1} ZEUS Collaboration:  S. Chekanov {\it et al}., Eur.
Phys. J. {\bf C21} (2001) 443.

\bibitem{ZEUS2} ZEUS Collaboration:  J. Breitweg {\it et al}., Phys.
Lett. {\bf B487} (2001) 53.

\bibitem{E665} M.~R. Adams {\it et al}., Phys. Rev. {\bf D54} (1996)
3006.

\bibitem{NLOimpact}
J. Bartels, S. Gieseke and C.~F. Qiao,
Phys. Rev. {\bf D63}, (2001) 056014,
[Erratum-ibid. {\bf D65}, (2002) 079902];\\
V.~S. Fadin, D.~Y. Ivanov and M.~I. Kotsky,
Phys.\ Atom.\ Nucl.\  {\bf 65}, (2002) 1513;\\
J. Bartels, S. Gieseke and A. Kyrieleis,
Phys.\ Rev.\  {\bf D65}, (2002) 014006;\\ 
J. Bartels, D. Colferai, S. Gieseke and A. Kyrieleis,
Phys.\ Rev.\ {\bf D66}, (2002) 094017;\\ 
J. Bartels, and A. Kyrieleis, {\tt hep-ph/0407051}.

\bibitem{NLOBFKL}
V.~S. Fadin and L.~N. Lipatov,
Phys.\ Lett.\ {\bf B429}, 127 (1998);\\
G. Camici and M. Ciafaloni, Phys. Lett. {\bf B430} (1998) 349.

\bibitem{Bialas1} 
A. Bialas, H. Navalet and R. Peschanski, Nucl. Phys. {\bf B603},
218, (2001).

\bibitem{Thorne}
R.~S. Thorne, Phys. Rev. {\bf D60} (1999) 054031.

\end{thebibliography}
\end{document}